\documentclass[12pt]{article} \usepackage{amssymb,latexsym}
\usepackage{amsmath}
 
\columnwidth\textwidth

\begin{document}

\newtheorem{df}{Definition} \newtheorem{thm}{Theorem} \newtheorem{lem}{Lemma}
\newtheorem{rl}{Rule}
\begin{titlepage}
 
\noindent
 
\begin{center} {\LARGE A new approach to quantum measurement problem: cluster
separability} \vspace{1cm}

P. H\'{a}j\'{\i}\v{c}ek \\ Institute for Theoretical Physics \\ University of
Bern \\ Sidlerstrasse 5, CH-3012 Bern, Switzerland \\ hajicek@itp.unibe.ch

\vspace{1cm}

March 2011 \\ Based on the talk at \\ Fifth International Workshop DICE2010,
\\ Castiglioncello (Tuscany), September 13--17, 2010 \vspace{1cm}
 
 
\vspace*{2cm}
 
\nopagebreak[4]
 
\begin{abstract} The paper describes a solution to the problem of quantum
measurement that has been proposed recently. The literal understanding of the
basic rule of quantum mechanics on identical particles violates the cluster
separation principle and so leads to difficulties. A proposal due to Peres of
how such difficulties could be removed is reformulated and extended. Cluster
separability leads to a locality requirement on observables and to the key
notion of separation status. Separation status of a microsystem is shown to
change in preparation and registration processes. The indispensability of
detectors plays an important role. Changes of separation status are
alterations of kinematic description rather than some parts of dynamical
trajectories and so more radical than 'collapse of the wave
function'. Textbook quantum mechanics does not provide any information of how
separation status changes run, hence new rules must be formulated. This
enables to satisfy the objectification requirement for registrations. To show
how the ideas work, a simplified model of registration apparatus is
constructed.
\end{abstract}

\end{center}

\end{titlepage}

\section{Introduction} It is well known that the state of quantum theory of
measurement is unsatisfactory. Let us mention just the excellent book
\cite{BLM} and more recent clearly written critical review \cite{ghirardi}.

In a series of papers \cite{PHJT,hajicek,hajicek2,survey}, we looked for
solutions of three interrelated problems of quantum mechanics, those of
measurement, of classical properties and of realist interpretation. The three
main ideas that we proposed were very different from what had ever been
published and it seems that we have been successful.

All previous attempts to formulate a realist interpretation of quantum
mechanics failed because everybody was looking for objective properties where
they could not be found, namely among values of observables. Quantum
observables are not objective, see e.g.\ \cite{BLM}. Their values can be
obtained only after an interaction of a quantum system with a registration
apparatus and so they are properties of a composite: microsystem +
macrosystem, and not of the microsystem alone.

Hence, we have rejected all these attempts and proposed (an extended
discussion is given in \cite{PHJT,survey}) instead the following:
\par \vspace*{.4cm} {\bf Basic ontological hypothesis of quantum mechanics} A
property is objective if its value is uniquely determined by a preparation
according to the rules of standard quantum mechanics. The 'value' is the value
of the mathematical expression that describes the property and it may be more
general than just a real number. No registration is necessary to establish
such a property but a correct registration cannot disprove its value; in many
cases, registrations can confirm the value.
\par \vspace*{.4cm}

Our interpretation of quantum mechanics differs further from Copenhagen
interpretation by assuming the universality of quantum mechanics: all physical
systems are quantum systems. Thus, classical systems are only a special kind
of quantum systems. Our theory of classical properties is given in
\cite{hajicek,survey}. It is based on the principle of maximum entropy rather
than on the popular but incorrect ideas about coherent states. For example, we
have introduced quantum states called maximum-entropy packets and shown them
to match trajectories of classical mechanics better than Gaussian wave
packets.

The present paper contains only a simplified version of some selected points
concerning the problem of measurement.

\section{Identical particles} We have to start with the standard theory of
identical particles. The basic rule is:
\begin{rl} Let system ${\mathcal S}$ consists of $N$ particles of the same
type, each particle having Hilbert space ${\mathbf H}$. Then, the Hilbert
space of ${\mathcal S}$ is the symmetrised tensor product of $N$ copies of
${\mathbf H}$, ${\mathbf H}_s^N$, for bosons and the anti-symmetrised one,
${\mathbf H}^N_a$, for fermions. States and observables of ${\mathcal S}$ are
then described by operators on the respective Hilbert space ${\mathbf H}_s^N$
or ${\mathbf H}^N_a$.
\end{rl} It is known that the literal understanding of Rule 1 leads to
problems (see e.g.\ \cite{peres} p 128). To explain the problems, let us
consider two experiments.
\par \vspace*{.4cm} \noindent {\bf Experiment I}: State $\psi(\vec{x}_1)$ of
particle ${\mathcal S}_1$ is prepared in our laboratory.
\par \vspace*{.4cm} \noindent {\bf Experiment II}: State $\psi(\vec{x}_1)$ is
prepared as in Experiment I and state $\phi(\vec{x}_2)$ of particle ${\mathcal
S}_2$ of the same type is prepared simultaneously in a remote laboratory.
\par \vspace*{.4cm} \noindent Now, our laboratory claims: the state of
${\mathcal S}_1$ is $\psi(\vec{x}_1)$ because this has been prepared according
to all rules of experimental art. An adherent of Rule 1 who knows about both
laboratories will, however, claim: the true preparation is a combination of
the two sub-preparations, one in each laboratory, and the state is
\begin{equation}\label{symstate} 2^{-1/2}\bigl(\psi(\vec{x}_1)\phi(\vec{x}_2)
\pm \phi(\vec{x}_1)\psi(\vec{x}_2)\bigr)\ ,
\end{equation} as Rule 1 requires.

Thus, it seems that the notions of preparation and of state are ambiguous. In
fact, Rule 1 requires to work with the state of {\em all} particles of the
same type in the universe and the uncertainty is much higher than that in the
above example.

Has this ambiguity any observable consequences? To answer this question, let
us first consider Experiment I supplemented by a registration corresponding to
the observable of ${\mathcal S}_1$ with kernel $a(\vec{x}_1;\vec{x}'_1)$, and
let the registration be made in our laboratory. Measurements of this kind lead
to average value
\begin{equation}\label{aver1} \int d^3x_1d^3x'_1\,
a(\vec{x}_1;\vec{x}'_1)\psi^*(\vec{x}_1)\psi(\vec{x}'_1)\ .
\end{equation} Second, perform Experiment II supplemented by the same
registration in our laboratory as above. The correct observable corresponding
to this registration, as required by Rule 1, now is:
\begin{equation}\label{symobs}
a(\vec{x}_1;\vec{x}'_1)\delta(\vec{x}_2-\vec{x}'_2) +
\delta(\vec{x}_1-\vec{x}'_1)a(\vec{x}_2;\vec{x}'_2)\ .
\end{equation} Such measurements lead to the average value defined by Eqs.\
(\ref{symstate}) and (\ref{symobs}):
\begin{equation}\label{aver2} \int d^3x_1d^3x'_1\,
a(\vec{x}_1;\vec{x}'_1)\psi^*(\vec{x}_1)\psi(\vec{x}'_1) + \int
d^3x_1d^3x'_1\, a(\vec{x}_1;\vec{x}'_1)\phi^*(\vec{x}_1)\phi(\vec{x}'_1)\ .
\end{equation} Expressions (\ref{aver1}) and (\ref{aver2}) differ by a term
that is equal to the average of observable $a(\vec{x}_1;\vec{x}'_1)$ in state
$\phi$. Hence, measurements on ${\mathcal S}_1$ are in general disturbed by
other identical systems.

\section{Cluster separability} Cluster separability principle is a kind of
locality assumption that has been fruitful in several branches of quantum
physics, see e.g.\ \cite{KP} and Chap.\ 4 of \cite{weinberg}. An application
of the principle to identical particles can be found for instance in
\cite{peres} p 128. We formulate it as follows:
\par \vspace*{.4cm} \noindent {\bf Cluster Separability Principle} No quantum
experiment with a system in a local laboratory is affected by the mere
presence of an identical system in remote parts of the universe.
\par \vspace*{.4cm} \noindent Peres also observed that the principle was in
general violated by results similar to those of the previous section, and he
suggested a solution. This solution will now be reformulated and extended.

We introduce an important locality property of observables (for generalisation
to composite systems, see \cite{survey}):
\begin{df} Let $D \subset {\mathbb R}^{3}$ be open. Operator with kernel
$a(\vec{x}_1;\vec{x}'_1)$ is \underline{$D$-local} if
$$
\int d^3x'_1\, a(\vec{x}_1;\vec{x}'_1) f(\vec{x}'_1) = \int d^3x_1\,
a(\vec{x}_1;\vec{x}'_1) f(\vec{x}_1) = 0\ ,
$$
for any test function $f$ that vanishes in $D$.
\end{df} Now assume for Experiment II that
\begin{enumerate}
\item our laboratory is inside open set $D \subset {\mathbb R}^{3}$,
\item $\text{supp}\,\phi \cap D = \emptyset$.
\end{enumerate} Then, clearly the second term in (\ref{aver2}) vanishes for
$D$-local observables and Eqs.\ (\ref{aver1}) and (\ref{aver2}) agree in this
case (for a more general theorem see \cite{survey}). This suggests our
strategy. In fact, what we shall do will formalise and explain the common
practice that serenely ignores all unknown identical systems and that is
intuitive, irrational but successful.
\begin{enumerate}
\item We introduce the key notion of our theory:
\begin{df} Let ${\mathcal S}$ be a particle and $D \subset {\mathbb R}^3$ an
open set satisfying the conditions:
\begin{itemize}
\item Registrations of any $D$-local observable ${\mathsf A}$ of ${\mathcal
S}$ lead to average $\langle \psi(\vec{x})|{\mathsf A}\psi(\vec{x})\rangle$
for all states $\psi(\vec{x})$ of ${\mathcal S}$.
\item ${\mathcal S}$ is prepared in state $\phi(\vec{x})$ such that
supp\,$\phi \cap D \neq \emptyset$.
\end{itemize} In such a case, we say that ${\mathcal S}$ has
\underline{separation status} $D$.
\end{df} (For generalisation to composite systems and non-vector states, see
\cite{survey}.) Thus, the registration is not disturbed by other states of
identical systems. This can be the case e.g.\ if wave functions of all other
identical systems vanish in $D$.
\item We assume: Any preparation of ${\mathcal S}$ must give it a non-trivial
separation status $D \neq \emptyset$. Then $D$-local observables are
individually registrable on ${\mathcal S}$ but \underline{only these are} if
${\mathcal S}$ has not separation status $D'$ such that $D$ is a proper subset
of $D'$. Indeed, registration of observables that are not $D$-local are then
disturbed by other identical systems. Now, Peres did not warn that standard
observables of quantum mechanics (position, momentum, energy, angular
momentum, spin...) were \underline{not} $D$-local with $D \neq {\mathbb R}^3$.
This problem can be removed by a construction of $D$-local observables that
are registered in real experiments and that resemble, in certain sense, the
standard ones \cite{survey}.
\item We list the cases in which Rule 1 does not hold:
\begin{rl} Let system $\mathcal S$ be prepared in a state described by state
operator ${\mathsf T}$ so that it has separation status $D \neq
\emptyset$. Then its state is ${\mathsf T}$ and its observables form algebra
${\mathbf A}[\mathcal S]_D$ of all $D$-local observables of $\mathcal S$.
\end{rl} Composition of such states and observables satisfy
\begin{rl} Let systems ${\mathcal S}_1$ and ${\mathcal S}_2$ be prepared in
states ${\mathsf T}_1$ and ${\mathsf T}_2$ with separation statuses $D_1$ and
$D_2$, respectively. Then system ${\mathcal S}_1 + {\mathcal S}_2$ has state
${\mathsf T}_1 \otimes {\mathsf T}_2$ and the algebra of its observables is
${\mathbf A}[{\mathcal S}_1]_{D_1} \otimes {\mathbf A}[{\mathcal S}_2]_{D_2}$.
\end{rl} Clearly, the preparation described in Rule 3 is a preparation of
${\mathcal S}_1 + {\mathcal S}_2$ in state ${\mathsf T}_1 \otimes {\mathsf
T}_2$ with separation status $D_1 \cup D_2$ only if $D_1 = D_2$.
\end{enumerate}

Now, cluster separability holds.

\section{Preparations and registrations} As usual, we assume that any
measurement on microsystems can be divided into a preparation and a
registration procedure. The preparation determines a state and the
registration gives a value of an observable. What do the above ideas imply for
preparation and registration?

As we have seen in point 2 of our strategy, any preparation must transfer
system ${\mathcal S}$ from a trivial into a non-trivial separation
status. Thus, the separation status changes during a preparation. What is the
relation of registrations to separation status change?

An important assumption of our theory of measurement is:
\begin{rl} Any registration apparatus for \underline{microsystems} must
contain at least one detector and every 'reading of a pointer value' (see
e.g.\ \cite{BLM}) is a signal from a detector.
\end{rl} Hence, during registration, system must enter the sensitive matter of
a detector. In this way, its non-trivial separation status changes into a
trivial one. More discussion of this important point is given in
\cite{hajicek2,survey}.

Let us give an example of separation status change. Let the states in Rule 3
be vector states of identical particles,
$$
{\mathsf T}_1 = \psi(\vec{x})\psi^*(\vec{x}')\ , \quad {\mathsf T}_2 =
\Psi(\vec{x}_1, \cdots, \vec{x}_N)\Psi^*(\vec{x}'_1, \cdots, \vec{x}'_N)\ .
$$
Then the state of ${\mathcal S}_1 + {\mathcal S}_2$ is
\begin{equation}\label{state1} \psi(\vec{x})\Psi (\vec{x}_1, \cdots,
\vec{x}_N)\ .
\end{equation} Let the time evolution leads to ${\mathcal S}_1$ entering $D_2$
and ${\mathcal S}_1 + {\mathcal S}_2$ having separation status $D_2$. Then the
state of ${\mathcal S}_1 + {\mathcal S}_2$ will be
\begin{equation}\label{state2} {\mathsf P}_{\mathbf S} \circ {\mathsf
P}^{(N+1)}_{s,a}\big(\psi(\vec{x})\Psi(\vec{x}_1, \cdots \vec{x}_N)\big)\ ,
\end{equation} where ${\mathsf P}^{(N+1)}_{s,a}$ denotes symmetrisation or
anti-symmetrisation of $N+1$ arguments and ${\mathsf P}_{\mathbf S}$ is the
projection to the unit sphere. The map
$$
{\mathsf P}_{\mathbf S} \circ {\mathsf P}^{(N+1)}_{s,a} : {\mathbf H} \otimes
{\mathbf H}^N_{s,a} \mapsto {\mathbf H}^{N+1}_{s,a}
$$
is a non-invertible and non-linear map between two different Hilbert
spaces. Second, in state (\ref{state1}), the observables that are registrable
on $\mathcal S$ form the algebra ${\mathbf A}[{\mathcal S}_1]_{D_1}$. In state
(\ref{state2}), the algebra would be ${\mathbf A}[{\mathcal S}_1]_\emptyset =
\emptyset$: there are no observables that would be registrable individually on
${\mathcal S}_1$. Thus, the set of observables changes from ${\mathbf
A}[{\mathcal S}_1]_{D_1}$ to ${\mathbf A}[{\mathcal S}_1]_\emptyset$.

In classical mechanics, the possible states of system ${\mathcal S}$ are all
positive normalised functions (distribution functions) on the phase space
${\mathbf P}$ and possible observables are all real function on ${\mathbf P}$
(at least, all such observables have definite averages on ${\mathcal S}$
independently of external circumstances). ${\mathbf P}$ is fixed and uniquely
associated with the system alone and forms the basis of this kinematic
description. Hence, transitions between different sets of observables similar
to those described above would be impossible in classical mechanics. They are
only enabled in quantum mechanics by the non-objective character of
observables: not only their values cannot be ascribed to microsystem
${\mathcal S}$ alone but some of them are not even registrable in principle
due to external conditions in which ${\mathcal S}$ is.

We assume that the quantum kinematics of a microsystem is defined
mathematically by the possible states represented by all positive normalised
(trace one) operators, and possible observables represented by some
self-adjoint operators, on the Hilbert space associated with the system. Then
the transitions of states and observables that go with changes of separation
status cannot be viewed as a part of a dynamical trajectory due to some new
version of the dynamics of ${\mathcal S}$, but as a change of its kinematic
description. Thus, although the change of separation status is similar to the
collapse of the wave function (the non-local character included), it is both
more radical and better understood.

What has been said up to now shows that textbook quantum mechanics is
incomplete in the following sense:
\begin{enumerate}
\item It accepts and knows only \underline{two} separation statuses:
\begin{enumerate}
\item that of isolated systems, $D = {\mathbb R}^3$, with the standard
operators as observables, and
\item that of a member of a system of identical particles, $D = \emptyset$,
with no registrable observables of its own.
\end{enumerate}
\item It provides \underline{no} rules for changes of separation status.
\end{enumerate} Our main idea is: Quantum mechanics must be supplemented by a
theory of general separation status and by new rules that govern processes in
which separation status changes. The new rules must not contradict the rest of
quantum mechanics and ought to agree with and explain observational facts.

\section{Extended understanding of separation status} Definition 2 leaves open
the question of what the nature of disturbances is that might prevent
registrations on ${\mathcal S}$. As yet, all examples of such a disturbance
had to do with entanglement of identical particles. In the light of Rule 4,
the absence of such entanglement need not be sufficient for ${\mathcal S}$ to
allow undisturbed registrations. To explain that, it is helpful to distinguish
two kinds of registration.

A {\em direct registration} first manipulates ${\mathcal S}$ by classical
fields and shields so that the prepared beam is split into spatially separated
beams, each of which associated with one eigenvalue of the registered
observable. Then, there is a set of detectors each of which can be hit by only
one beam. Hence, ${\mathcal S}$ must be separated from other microsystems, not
only of the same type, so hat it is ${\mathcal S}$ that is available to those
manipulation by fields and shields and has sufficient kinetic energy to excite
a detector.

An {\em indirect registration}, such as scattering or QND measurement (see
e.g.\ \cite{bragin} and \cite{peres} p 400), lets ${\mathcal S}$ interact with
another microsystem ${\mathcal S}'_1$ and it is only ${\mathcal S}'_1$ that is
then subject to a direct registration. For the measurement to be QND, several
further conditions must be satisfied, but this does not concern us here. After
a QND procedure, ${\mathcal S}$ remains available to another one: another
system ${\mathcal S}'_2$ of the same type as ${\mathcal S}'_1$ interacts with
${\mathcal S}$ and is then directly registered etc. Information given by the
detectors of the direct registrations reveals also something about ${\mathcal
S}$. Thus, detectors are necessary for indirect registrations, too, and
${\mathcal S}$ must be suitably separated from other microsystems, not only of
the same type, so that we can be sure that it was ${\mathcal S}$ that
interacted with ${\mathcal S}'_1$, ${\mathcal S}'_2$ etc.

There are interesting consequences for macroscopic systems. In general, a
macroscopic system ${\mathcal A}$ contains very many different
particles. Consider the observable with kernel $a(\vec{x}_k;\vec{x}'_k)$ that
concerns particle ${\mathcal S}$ of ${\mathcal A}$. Suppose that there would
be an apparatus ${\mathcal B}$ associated with registration of
$a(\vec{x}_k;\vec{x}'_k)$ on ${\mathcal S}$, if ${\mathcal S}$ were prepared
individually. Then, the apparatus ${\mathcal B}$ cannot be applied to
${\mathcal A}$ to register $a(\vec{x}_k;\vec{x}'_k)$ because ${\mathcal S}$ is
not suitably separated.

For example, in a direct registration, readings of ${\mathcal B}$ are signals
of a detector that registers ${\mathcal S}$ and ${\mathcal S}$ must be
isolated to be manipulable, have sufficient kinetic energy, etc. Hence, to
register $a(\vec{x}_k;\vec{x}'_k)$, we need a method that makes measurements
directly on ${\mathcal A}$ and which is, therefore, different from apparatus
${\mathcal B}$.

Consider a scattering as an example of indirect registration. Let ${\mathcal
A}$ be a crystal. By scattering $X$-rays off it, relative positions of its
ions can be recognised. But rather than a position of an individual ion it is
a space dependence of the average density due to all ions. In general,
scattering of a microsystem ${\mathcal S}'$ off a macrosystem ${\mathcal A}$
can be determined in terms of potential $V_k(\vec{x}, \vec{x}_k)$ that
describes the interaction between ${\mathcal S}'$ and $k$-th microscopic
subsystem ${\mathcal S}_k$ of ${\mathcal A}$ so that the whole interaction
Hamiltonian is the sum
$$
\sum_k V_k(\vec{x}, \vec{x}_k)\ .
$$
Hence, the scattering yields only information on a sum of over all subsystems
${\mathcal S}_k$ of ${\mathcal A}$ that can interact with ${\mathcal S}'$ and
${\mathcal S}_k$'s need not be identical with ${\mathcal S}'$. Although
ingenious potentials can be invented, nature provides only a small number of
potentials.

Another example is the kinetic energy of ${\mathcal S}$. Again, this
observable cannot be measured by the method kinetic energy is measured on
individual systems of type ${\mathcal S}$ (proportional counter, scattering
off a crystal, etc.). In the case that ${\mathcal A}$ is in a state of
thermodynamic equilibrium, the average kinetic energy can be calculated from
the temperature. Hence, a viable method to measure the average is to measure
the temperature of ${\mathcal A}$. Again, this is a special case that works
only under specific conditions. Further examples could be other additive
quantities, such as momentum and angular momentum. Average total values of
these quantities can be measured directly on ${\mathcal A}$. We notice that
only the averages of some observables with rather large variances can be
observed in these cases. It is impossible to obtain the single eigenvalues of
these observables as results of registration (for an example, see Ref.\
\cite{peres} p 181).

Thus, the qualitative difference between macroscopic and microscopic systems
is due to quantum properties of macroscopic systems rather than to some kind
of limits on the validity of quantum mechanics for them.

\section{ Beltrametti-Cassinelli-Lahti model of quantum measurement} We shall
proceed from account \cite{BLM} (p 38) of model \cite{belt}. Let a discrete
observable ${\mathsf O}$ of microsystem ${\mathcal S}$ with Hilbert space
${\mathbf H}$ is registered. Let $o_k$ be the eigenvalues and $\{\phi_{kj}\}$
be the complete orthonormal set of eigenvectors of ${\mathsf O}$,
$$
{\mathsf O}\phi_{kj} = o_k \phi_{kj}\ .
$$
We assume that $ k = 1,\cdots,N$ so that there is only a finite number of
different eigenvalues $o_k$. This is justified by the fact that no real
registration apparatus can distinguish all elements of an infinite set from
each other. It can therefore measure only a function of an observable that
maps its spectrum onto a finite set of real numbers. Our observable ${\mathsf
O}$ is such a function.

Let the registration apparatus be a quantum macrosystem ${\mathcal A}$ with
Hilbert space ${\mathbf H}_{\mathcal A}$. Let ${\mathcal S}$ and ${\mathcal
A}$ be prepared in some independent initial states and then interact for a
finite time by coupling ${\mathsf U}$, where ${\mathsf U}$ is a unitary
transformation on ${\mathcal H}_{\mathcal S}\otimes {\mathcal H}_{\mathcal
A}$. Then, a theorem has been shown in \cite{belt}:
\begin{thm} For any initial vector state $\psi$ of ${\mathcal A}$, there is a
set $\{\varphi_{kl}\}$ of unit vectors in ${\mathcal H}_{\mathcal S}$
satisfying the orthogonality conditions
$$
\langle \varphi_{kl}|\varphi_{kj}\rangle = \delta_{lj}
$$
such that ${\mathsf U}$ is a unitary extension of the map
\begin{equation}\label{unitar} \phi_{kl}\otimes \psi \mapsto
\varphi_{kl}\otimes \psi_k\ ,
\end{equation} where $\{\psi_k\}$ is a set of $N$ orthonormal vectors in
${\mathbf H}_{\mathcal A}$.
\end{thm}

One assumes that states $\psi_k$ are uniquely associated with what will be
read on the apparatus after the measurement. Then, an important requirement is
that the apparatus is in one of the states $|\psi_k\rangle\langle\psi_k|$
after each individual registration. This is called {\em objectification
requirement}.

Suppose that the initial state of ${\mathcal S}$ is an eigenstate, ${\mathsf
T} =|\phi_{kl}\rangle\langle\phi_{kl}|$, with the eigenvalue $o_k$. Then, Eq.\
(\ref{unitar}) implies that the final state of apparatus ${\mathcal A}$ is
$|\psi_k\rangle\langle\psi_k|$, and the registration does lead to a definite
result. However, suppose next that the initial state is an arbitrary vector
state, ${\mathsf T} =|\phi\rangle\langle\phi|$. Decomposing $\phi$ into the
eigenstates,
$$
\phi = \sum_{kl} c_{kl}\phi_{kl}\ ,
$$
we obtain from Eq.\ (\ref{unitar})
\begin{equation}\label{finalSA} {\mathsf U} (\phi \otimes \psi) = \sum_k
\sqrt{p_k}\Phi_k\otimes \psi_k\ ,
\end{equation} where
\begin{equation}\label{Phik} \Phi_k = \frac{\sum_l
c_{kl}\varphi_{kl}}{\sqrt{\langle \sum_l c_{kl}\varphi_{kl}|\sum_j
c_{kj}\varphi_{kj}\rangle}}
\end{equation} and
$$
p_k = \left\langle \sum_l c_{kl}\varphi_{kl}\Biggm|\sum_j
c_{kj}\varphi_{kj}\right\rangle
$$
is the probability that a registration of ${\mathsf O}$ performed on vector
state $\phi$ gives the value $o_k$.

The final state of apparatus ${\mathcal A}$ is the partial trace over
${\mathcal S}$:
\begin{equation}\label{finalA} tr_{\mathcal S}[{\mathsf U}({\mathsf T}\otimes
{\mathsf T}_{\mathcal A}){\mathsf U}^\dagger] = \sum_{kl}
\sqrt{p_k}\sqrt{p_l}\langle\Phi_k|\Phi_l\rangle |\psi_k\rangle\langle\psi_l|\
.
\end{equation} If the objectification requirement is to be satisfied, two
condition must be met:
\begin{description}
\item[(A)] The final state of the apparatus must the convex combination of the
form
\begin{equation}\label{gemengA} tr_{\mathcal S}[{\mathsf U}({\mathsf T}\otimes
{\mathsf T}_{\mathcal A}){\mathsf U}^\dagger] = \sum_j
p_j|\psi_j\rangle\langle\psi_j|\ .
\end{equation}
\item[(B)] The right-hand side of Eq.\ (\ref{gemengA}) must be the {\em
gemenge structure} of the state.
\end{description}

The notion of gemenge will play an important role. The term has been
introduced in Ref.\ \cite{BLM}, some authors use the term 'proper mixture' or
'direct mixture'. The crucial point is that the convex decomposition
\begin{equation}\label{defgem} {\mathsf T} = \sum_{k=1}^n w_k {\mathsf T}_k
\end{equation} of any state ${\mathsf T}$ can be a gemenge only if its
preparation procedure ${\mathbf P}({\mathsf T})$ is a random mixture with
rates (frequencies) $w_k$ of preparations ${\mathbf P}({\mathsf T}_k)$, where
each ${\mathbf P}({\mathsf T}_k)$ is some preparation procedure for ${\mathsf
T}_k$, $k = 1,\cdots,n$. The preparation mixture can be done by humans or
result from some process in nature.

Thus, gemenge concerns a physical property of preparation rather than any
mathematical characteristic of the right-hand side of Eq.\ (\ref{defgem})
(such as ${\mathsf T}_k$ being vector states or being mutually orthogonal,
etc). From the mathematical point of view, many different convex
decompositions of a general state ${\mathsf T}$ may exist. A state is
'extremal' if it cannot be written as a non-trivial convex
combination. Extremal states are described by projections onto one-dimensional
subspaces of the Hilbert space. A preparation of non-extreme state ${\mathsf
T}$ selects only some of the mathematically possible convex decompositions of
${\mathsf T}$.

A random mixture of preparations is not uniquely determined by the preparation
process. It can be coarsened or refined i.e. some of ${\mathbf P}({\mathsf
T}_k)$ can be combined into one preparation procedure or ${\mathbf P}({\mathsf
T}_k)$ for some $k$ can itself be a random mixture of other preparations.
\begin{df} The finest convex decomposition of state ${\mathsf T}$ defined by
its preparation as gemenge is called {\em gemenge structure} of ${\mathsf T}$.
\end{df} Thus, gemenge structure of ${\mathsf T}$ is uniquely determined by
its preparation.

It may be advantageous to distinguish the mathematical convex combination of
states from its gemenge structure by writing the sum in Eq.\ (\ref{defgem}) as
follows
\begin{equation}\label{defgem'} {\mathsf T} =
\left(\sum_{k=1}^n\right)_{\text{gs}} w_k {\mathsf T}_k
\end{equation} in the case that the right-hand side is a gemenge structure of
${\mathsf T}$.

The properties that follow directly from the definition of gemenge structure
and that will be needed later are described by the following theorem.
\begin{thm}
\begin{enumerate}
\item Gemenge structure is preserved by unitary dynamics,
$$
{\mathsf U}\left[\left(\sum_k\right)_{\text{gs}} w_k {\mathsf
T}_k\right]{\mathsf U}^\dagger = \left(\sum_k\right)_{\text{gs}} w_k {\mathsf
U}{\mathsf T}_k{\mathsf U}^\dagger\ :
$$
if the sum on the left-hand side describes a gemenge structure of ${\mathsf
T}$, then the gemenge structure of its evolution is described by the sum on
the right-hand side.
\item In the following sense, gemenge structure is also preserved by
composition of systems. Let ${\mathsf T}$ be a state of a composite system
${\mathcal S} + {\mathcal S}'$. The necessary and sufficient condition for the
partial trace over ${\mathcal S}'$ to have the gemenge structure described by
$$
tr_{{\mathcal S}'}[{\mathsf T}] = \left(\sum_k\right)_{\text{gs}} w_k {\mathsf
T}_k
$$
is that ${\mathsf T}$ itself has gemenge structure described by
$$
{\mathsf T} = \left(\sum_k\right)_{\text{gs}} w_k {\mathsf T}_k \otimes
{\mathsf T}'_k\ ,
$$
where ${\mathsf T}'_k$ are some states of ${\mathcal S}'$.
\end{enumerate}
\end{thm}

All these ideas on gemenges seem to be well known. Now, an important new point
will be added: the Basic Ontological Hypothesis of Quantum Mechanics (see
Introduction). It leads to a new meaning of gemenge structure: any individual
system prepared in the state (\ref{defgem'}) is objectively in one of the
states ${\mathsf T}_k$, because each of the systems has been prepared by one
of the preparations ${\mathbf P}({\mathsf T}_k)$, and the probability that
${\mathbf P}({\mathsf T}_k)$ has been used is $w_k$.

It follows that the contents of both points {\bf (A)} and {\bf (B)} can be
concisely written as
$$
tr_{\mathcal S}[{\mathsf U}({\mathsf T}\otimes {\mathsf T}_{\mathcal
A}){\mathsf U}^\dagger] = \left(\sum_j\right)_{\text{gs}}
p_j|\psi_j\rangle\langle\psi_j|\ .
$$

Beltrametti-Cassinelli-Lahti model does not satisfy the objectification
criterion because the end state ${\mathsf T}\otimes {\mathsf T}_{\mathcal A}$
of the system is ${\mathsf U} (\phi \otimes \psi)$ (Eq.\ (\ref{finalSA})),
which is a vector state and can therefore have only a trivial gemenge
structure. Then, Point 2 of Theorem 2 implies that this is not compatible with
state $tr_{\mathcal S}[{\mathsf U}({\mathsf T}\otimes {\mathsf T}_{\mathcal
A}){\mathsf U}^\dagger]$ being a non-trivial gemenge.

\section{Model of registration apparatus} To show, how our ideas on
measurement work, we have constructed a model of registration apparatus
\cite{hajicek2,survey}. It starts by extending and modifying
Beltrametti-Cassinelli-Lahti model by additional assumptions:
\begin{enumerate}
\item Particle ${\mathcal S}$ is prepared in state $\phi$ and separation
status $D$.
\item ${\mathcal A}$ is an array of $N$ detectors ${\mathcal A}_k$ and each
detector has separation status $D_k$ where all $D_k$ are mutually disjoint and
disjoint from $D$.
\item The support of state $\varphi_{kl}$ is $D_k$ for all $k$ and $l$. It
follows then that,
\begin{equation}\label{endorth} \langle \varphi_{kl}|\varphi_{mn}\rangle =
\delta_{km}\delta_{ln}\quad k\leq N, m\leq N\ .
\end{equation}
\item Each detector is composite, ${\mathcal A}_k = {\mathcal A}'_k +
{\mathcal S}'_k$, where ${\mathcal S}'_k$ consists of all particles of
${\mathcal A}_k$ identical with ${\mathcal S}$. Let the number of particles in
${\mathcal S}'_k$ be $M_k$ and let the state of ${\mathcal S}'_k$ be ${\mathsf
T}_k$. Systems ${\mathcal S}'_k$ either are 'natural' parts of ${\mathcal
A}_k$'s or they are created quickly after the start of the measurement by
pollution.
\item ${\mathcal A}'_k$ is the part of the sensitive matter interacting with
${\mathcal S}$ so that $\psi_k$ is a state of ${\mathcal A}'_k$.
\item State (\ref{finalSA}) is not of an end state of ${\mathcal S} +
\sum_k{\mathcal A}_k'$ but a choice of an intermediate state before an
amplification procedure.
\end{enumerate}.

The existence of systems ${\mathcal S}'_k$ leads to a change of the separation
status of ${\mathcal S}$ after ${\mathcal S}$ enters a detector. There are
several mathematical possibilities for the choice of the intermediate
state. One of the possibilities is:
\begin{rl} The intermediate state of ${\mathcal S} + \sum_k{\mathcal S}'_k +
\sum_k{\mathcal A}'_k$ (before amplification) is
\begin{equation}\label{trig} {\mathsf T}_{\text{int}} =
\left(\sum_k\right)_{\text{gs}} p_k \nu^2_k {\mathsf T}_1 \otimes \cdots
\otimes {\mathsf T}_{k-1} \otimes {\mathsf W}_{kk} \otimes {\mathsf T}_{k+1}
\otimes \cdots \otimes {\mathsf T}_N \otimes |\psi_k\rangle \langle \psi_k|\ .
\end{equation}
\end{rl} Choices:
\begin{enumerate}
\item The term
$$
{\mathsf W}_{kk} = {\mathsf P}^{(M_k+1)}_{s,a} (|\varphi^1_k\rangle \langle
\varphi^1_k|\otimes {\mathsf T}_k){\mathsf P}^{(M_k+1)}_{s,a}
$$
expresses our choice of symmetrisation or anti-symmetrisation and $\nu_k^2$ is
a normalisation factor that makes ${\mathsf W}_{kk}$ to a state (action of
${\mathsf P}_{\mathbf S}$).
\item Some correlations have been erased so that the state operator ${\mathsf
T}_{\text{int}}$ is diagonal.
\item The convex decomposition (\ref{trig}) is postulated to be the gemenge
structure of ${\mathsf T}_{\text{int}}$.
\end{enumerate} The choices 2. and 3. are dictated by the objectification
requirement which can be regarded as an experimental fact.

Finally, we can show that the new rule cannot be disproved by measuring the
'erased' correlations if standard quantum mechanical rules on jointly
measurable observables hold \cite{hajicek2,survey}.

\section{Conclusion} We have improved understanding of the theory of identical
particles and removed a disorder in textbook quantum mechanics. This has lead
to the notion of separation status. We have discovered the crucial role that
detectors have in the theory. Then, both preparations and registrations
include changes of separation status, which are changes of kinematic
description and so even more radical than 'collapse of the wave function'. The
theoretical freedom in changes of separation status has been used to satisfy
objectification requirement. Thus, a deep revision of quantum theory of
measurement results that has been derived with the help of standard principles
of quantum mechanics from analysis of real measurement processes.

As yet, our results are limited to non-relativistic quantum mechanics and to
measurements performed on microsystems. Moreover, only special kind of
measurements have been considered, registrations of discrete observables with
the help of unitary measurement couplings, the definition assumptions of
Beltrametti-Cassinelli-Lahti model. Much work remains to be done.

\subsection*{Acknowledgements} The author is indebted to \v{S}tefan
J\'{a}no\v{s} and Ji\v{r}\'{\i} Tolar for important suggestions.

\end{document}